\documentclass[12pt,fleqn]{article}
\usepackage{amsmath,amsfonts,latexsym}
\usepackage[dvips]{graphics}

\usepackage{amsfonts,amssymb,cite}
\usepackage{graphicx}

\def\onecol{\onecolumn \mathindent 2em}

\def\noi{\noindent}

\newcommand{\Title}[1]{\noi {{\Large\bf #1}}\\[1ex]}

\def\Aunames#1{\noi{\bf #1}}
\def\au#1{${}^{#1}$}
\def\Addresses#1{\medskip\noi \protect
	\begin{description}\itemsep -3pt {\it #1} \end{description}}
\def\adr#1#2{\item[${}^{#1}$]{\it #2}}

\newcommand{\Abstract}[1]{\vskip 2mm \begin{center}
        \parbox{16.4cm}{\small\noi #1} \end{center}\medskip}

\def\email#1#2{\footnotetext[#1]{e-mail: #2}\addtocounter{footnote}{1}}

\def\nqq{\hspace*{-2em}}
\def\nhq{\hspace*{-0.5em}}

\def\qq{\qquad}
\def\cm{\hspace*{1cm}}

\def\wide{\mbox{$\dst\vphantom{\int}$}}

\usepackage{color}

\def\Funding#1{\subsection*{Funding} #1}

\def\Jl#1#2{#1 {\bf #2},\ }

\def\ApJ#1 {\Jl{Astroph. J.}{#1}}
\def\CQG#1 {\Jl{Class. Quantum Grav.}{#1}}
\def\DAN#1 {\Jl{Dokl. AN SSSR}{#1}}
\def\GC#1 {\Jl{Grav. Cosmol.}{#1}}
\def\GRG#1 {\Jl{Gen. Rel. Grav.}{#1}}
\def\IJMPD#1 {\Jl{Int. J. Mod. Phys. D}{#1}}
\def\JETF#1 {\Jl{Zh. Eksp. Teor. Fiz.}{#1}}
\def\JETP#1 {\Jl{Sov. Phys. JETP}{#1}}
\def\JHEP#1 {\Jl{JHEP}{#1}}
\def\JMP#1 {\Jl{J. Math. Phys.}{#1}}
\def\NPB#1 {\Jl{Nucl. Phys. B}{#1}}
\def\NP#1 {\Jl{Nucl. Phys.}{#1}}
\def\PLA#1 {\Jl{Phys. Lett. A}{#1}}
\def\PLB#1 {\Jl{Phys. Lett. B}{#1}}
\def\PRD#1 {\Jl{Phys. Rev. D}{#1}}
\def\PRL#1 {\Jl{Phys. Rev. Lett.}{#1}}

\def\al{&\nhq}
\def\lal{&&\nqq {}}
\def\eq{Eq.\,}
\def\eqs{Eqs.\,}
\def\beq{\begin{equation}}
\def\eeq{\end{equation}}
\def\besub{\begin{subequations}}
\def\esub{\end{subequations}}
\def\bear{\begin{eqnarray}}
\def\bearr{\begin{eqnarray} \lal}
\def\ear{\end{eqnarray}}
\def\earn{\nonumber \end{eqnarray}}

\def\nnn{\nonumber\\ \lal }

\def\yy{\\[5pt] {}}
\def\yyy{\\[5pt] \lal }
\def\eql{\al =\al}

\def\dst{\displaystyle}

\def\e{{\,\rm e}}
\def\d{\partial}

\def\sign{\mathop{\rm sign}\nolimits}

\def\const{{\rm const}}
\def\eps{\varepsilon}

\def\then{\ \Rightarrow\ }

\def\eqn#1{\eq\eqref{#1}}
\def\rf{\eqref}

\def\M{{\mathbb M}}
\def\ME{\mbox{$\M_{\rm E}$}}
\def\MJ{\mbox{$\M_{\rm J}$}}

\def\R{{\mathbb R}}
\def\Z{{\mathbb Z}}

\def\og{{\overline g}}

\def\oR{{\overline R}}
\def\oom{{\overline \omega}}
\def\oq{{\overline q}{}}

\def\df{\delta\phi}
\def\da{\delta\alpha}
\def\db{\delta\beta}
\def\dg{\delta\gamma}
\def\dpsi{\delta\psi}

\def\kappa{\varkappa}
\def\Geff{G_{\rm eff}}
\def\Veff{V_{\rm eff}}

\def\GR{general relativity}
\def\pb{perturbation}
\def\pbs{perturbations}
\def\sph{spherically symmetric}
\def\ssph{static, spherically symmetric}
\def\asflat{asymptotically flat}
\def\wh{wormhole}
\def\whs{wormholes}

\def\bhs{black holes}

\def\icons{integration constants}

\def\emag{electromagnetic}

\def\Scw{Schwarz\-schild}
\def\Schr{Schr\"o\-din\-ger}

\def\mn{_{\mu\nu}}
\def\MN{^{\mu\nu}}

\usepackage{color}

\begin{document}
\thispagestyle{empty}
\onecol
\vspace*{10mm}

\Title{On the stability of exceptional Brans-Dicke wormholes}

\Aunames{Kirill A. Bronnikov,\au{a,b,c,1} Sergei V. Bolokhov,\au{b,2} Milena V. Skvortsova,\au{b,3}\\
			Rustam Ibadov,\au{d,4} and Feruza Y. Shaymanova\au{d,f,5} }
	
\Addresses{\small
\adr a	{Center of Gravitation and Fundamental Metrology, Rostest, 
		Ozyornaya ul. 46, Moscow 119361, Russia}
\adr b	{Institute of Gravitation and Cosmology, RUDN University, 
		ul. Miklukho-Maklaya 6, Moscow 117198, Russia}
\adr c 	{National Research Nuclear University ``MEPhI'', 
		Kashirskoe sh. 31, Moscow 115409, Russia}
\adr d {Department of Theoretical Physics and Computer Science, Samarkand State University, 
		Samarkand 140104, Uzbekistan}
\adr e {Karshi State  University, Kochabog street, Karshi city 180103, Uzbekistan}
		}

\Abstract
   {In our previous papers we have analyzed the stability of vacuum and electrovacuum static,
   spherically symmetric space-times in the framework of the Bergmann-Wagoner-Nordtvedt 
   class of scalar-tensor theories (STT) of gravity. In the present paper, we continue this study
   by examining the stability of exceptional solutions of the Brans-Dicke theory with the 
   coupling constant $\omega =0$ that were not covered in the previous studies. Such solutions 
   describe neutral or charged wormholes and involve a conformal continuation: the standard
   conformal transformation maps the whole Einstein-frame manifold \ME\ to only a part 
   of the Jordan-frame manifold \MJ, which has to be continued beyond the emerging regular
   boundary S, and the new region maps to another manifold $\ME_-$. The metric in \MJ\
   is symmetric with respect to S only if the charge $q$ is zero. Our stability study concerns 
   radial (monopole) perturbations, and it is shown that the wormhole is stable if $q \ne 0$
   and unstable only in the symmetric case $q=0$.
    }

\email 1 {kb20@yandex.ru} 
\email 2 {boloh@rambler.ru} 
\email 3 {milenas577@mail.ru} 
\email 4 {ibrustam@mail.ru}
\email 5 {yusupovnafiruz89@gmail.com}

\section{Introduction}

  In our previous papers \cite{we23, we24} we discussed the stability of \ssph\ vacuum and 
  electrovacuum solutions in scalar-tensor theories (STT) of gravity from the 
  Bergmann-Wagoner-Nordtvedt class \cite{STT1, STT2, STT3}. As its special cases, we 
  considered the Brans-Dicke (BD) theory \cite{BD-STT}, Barker's \cite{barker} and 
  Schwinger's \cite{schwg, bruk94} theories, and the case of general relativity (GR) with 
  nonminimally coupled scalar fields and an arbitrary nonminimality parameter $\xi$. In this class 
  of theories, physically relevant solutions within the original formulation of the theory (the Jordan 
  frame) are connected with those of  GR (the Einstein frame) sourced by a minimally coupled 
  scalar field and an \emag\ field \cite{fisher, penney} by a conformal transformation \cite{STT2},
  where the conformal factor depends on the particular theory.  
  This allowed us to obtain stability conclusions for the solutions in question, which contained 
  naked singularities \cite{br73} and were directly mapped to the GR solutions. 

  In this paper, we consider the stability of an exceptional solution of the BD  theory 
  that was not included in \cite{we23, we24}. This solution describes a
  charged or neutral \wh\ and corresponds to the special case $\omega =0$ of the BD theory
  under some relationship between the \icons\ at which the manifold \MJ\ of Jordan's frame 
  is obtained from its counterpart \ME\ in the Einstein frame with the aid of conformal
  continuation, and in fact each of two halves of this \wh\ configuration is conformal to its own
  Einstein-frame manifold, $\ME_+$ and $\ME_-$. Therefore, the stability of such a \wh\ 
  requires a special study, combining the analysis in these manifolds.
       
  Let us recall that an important feature of gravitating configurations involving scalar fields 
  is that their \pbs\ contain a monopole degree of freedom which most likely leads to an instability 
  of isolated field distributions. This happens because in the wave equations for \pbs\ of different
  multipolarity $\ell$, the corresponding effective potentials always contain a ``centrifugal barrier term''
  having the form $\ell(\ell+1)/r^2$. Meanwhile, when solving the boundary-value problems with
  these equations, such positive barriers can only increase the eigenvalues that in such problems 
  have the physical meaning of squared frequencies $\Omega^2$ of allowed \pbs. Thus possible 
  eigenvalues $\Omega^2 \leq 0$, which correspond to exponentially (if $\Omega^2 < 0$)
  or linearly (if $\Omega^2 = 0$) growing \pbs\ most likely emerge at the smallest existing 
  multipolarity $\ell$, which can be zero in the presence of scalar fields. And indeed, many
  configurations with scalar fields, including \bhs, \whs\ and boson stars, have turned out to be 
  unstable under such monopole \pbs, see, 
  e.g.,\,\cite{kb-hod, sarb1, sarb2, stab11, stab12, stab18, brito23, tugai25} and references therein.
  
  In \cite{we23, we24}, our study was restricted to STT with a canonical scalar field.
  This choice was not only motivated by a more evident physical relevance of canonical fields as 
  compared to ,phantom ones, bu talso by the fact that STT solutions with phantom scalars are
  conformally related to other branches of the GR solutions, their properties are quite different
  from those with canonical fields, 
  and in general require different methods of stability investigations because such solutions 
  generically describe \whs\ or at least contain \wh\ throats, and they in turn require Darboux 
  transformations able to regularize the \pb\ potentials \cite{sarb1, sarb2, stab11, stab12, stab18}. 
  Unlike that, with canonical scalars, \whs\ with correct asymptotic behaviors can only emerge in 
  exceptional cases due to conformal continuations \cite{br73, kb-CC2, br-star07}, such that the whole
  manifold \ME\ maps to a part of \MJ, and it is thus necessary to extend \MJ\ to a new region 
  with, generally, a negative effective gravitational constant $\Geff$ \cite{kb-CC2, br-star07, skvo10}. 
  Examples of such \whs, both neutral and charged, were previously found among solutions of GR 
  with nonminimally coupled scalars (considering this theory as a special case of STT) 
  \cite{br73, kb-CC2, bar-vis2, stepan1}  and shown to be unstable \cite{stepan1, stepan2}. 
  This paper is devoted to a study of one more such case, the BD theory with $\omega=0$,
  in which the existence of vacuum \whs\ was probably first noticed in \cite{br96}, while
  their electrovacuum extensions seem to be studied here for the first time. It turns out that 
  such vacuum BD \whs\ are $\Z_2$-symmetric with respect to their throats while the electrovacuum
  ones are asymmetric, and this strongly affects their stability properties. 
  
  The point is that the stability study consists in solving boundary-value problems for the \pb\
  equations, formulated in the Einstein-frame manifolds $\ME_+$ and $\ME_-$. Meanwhile, since 
  \MJ\ is a unique smooth manifold, its physically relevant \pbs\ can only exist if there are common
  eigenvalues of the two boundary-value problems. To cause an instability, such \pbs\ must grow  
  with time exponentially or linearly. In the case under consideration, this happens 
  only in the symmetric case $q=0$, therefore, only such \whs\ turn out to be unstable.
  
  The paper is organized as follows. Section 2 contains a derivation and a description of the \wh\
  solution to be studied. In Section 3 we present the equations for \sph\ \pbs\ of these \whs,
  to be used in Section 4 where we discuss the boundary conditions for perturbations and 
  describe a numerical study leading to our stability inferences. Section 5 is a conclusion.
  
   This paper may be considered as a natural addition to \cite{we23, we24}, but its content 
   and results illustrates an interesting opportunity of the existence and stability of new objects
   related to conformal continuations from GR to other metric theories of gravity.
  
\section{The Brans-Dicke theory: Electrovacuum solutions}

  We will deal here with the Brans-Dicke theory \cite{BD-STT} described by the action 
\bearr   \label{S_J}
             S_{\rm BD} = \frac 1{16\pi} \int \sqrt{-g} d^4 x
		             \Big[\phi R + \frac {\omega}{\phi}g\MN \phi_{,\mu} \phi_{,\nu} 
  			           - 2 U(\phi) + L_m\Big],
\ear    
  where $R$ is the scalar curvature, $g = \det(g\mn)$, $\omega \ne -3/2$ is the Brans-Dicke 
  coupling constant, $U(\phi)$ an arbitrary function (self-interaction potential of the $\phi$ field), 
  and $L_m$ the Lagrangian of any nongravitational matter. 
  This action corresponds to Jordan's (conformal) frame specified in pseudo-Riemannian 
  space-time \MJ\ with the metric $g\mn$. The conformal mapping \cite{BD-STT, STT2}
  with the accompanying scalar field substitution
\bearr              \label{map}
		g\mn = \og\mn/\phi, 
\nnn		
		\phi = \e^{\sqrt 2 (\psi-\psi_0)/\oom}, \quad\
		\psi_0=\const, \quad\   \oom = \sqrt{|\omega + 3/2|},
\ear 
  transforms the theory to the Einstein frame, specified in space-time \ME\ with 
  the metric $\og\mn$, in which the action becomes that of \GR\ with the 
  minimally coupled scalar field $\psi$,
\bearr   \label{S_E}
             S_{\rm STT} = \frac 1{16\pi} \int \sqrt{-\og} d^4 x
		             \Big[\oR + 2 \eps \og\MN \psi_{,\mu} \psi_{,\nu} 
  			           - 2 U(\phi)/\phi^2 + L_m/\phi^2\Big],
\ear      
  where bars mark quantities obtained from or with $\og\mn$, while
  $\eps = \sign (\omega + 3/2)$ distinguishes a canonical field $\psi$ ($\eps = +1$)
  from a phantom one ($\eps =-1$).
  
  Evidently, if a solution $(\og\mn, \psi)$ to the field equations is known in $\ME$, its 
  counterpart in $\MJ$ is also known, with the metric  
\beq             \label{ds_J}
		  ds_J^2 = g\mn dx^\mu dx^\nu = \frac 1{\phi} ds_E^2 
		  \								 = \frac 1{\phi}\og\mn dx^\mu dx^\nu
\eeq   
  and the $\phi$ field found according to \rf{map}. Here and further on the indices $E$ and $J$ 
  will be used to mark quantities belonging to \ME\ and \MJ, respectively.

 \subsection{Scalar-electrovacuum solution in \ME.}

  We will consider \sph\ solutions of the STT \rf{S_J} assuming $U(\phi) =0$ and matter in the 
  form of the Maxwell \emag\ field with $L_m = -F\mn F\MN$,  
\bear                                                  \label{S-GR}
     S _E =\frac 1{16\pi} \int {\sqrt{-\og}} \Big(\oR 
	         + 2\eps \og^{\alpha\beta}\psi_{,\alpha}\psi_{,\beta}- F\mn F\MN\Big),
\ear
  and, as usual, $F\mn = \d_\mu A_\nu - \d_\nu A_\mu$; for convenience, the gravitational 
  constant $G$ is absorbed in the definitions of $\psi$ and $F\mn$.

  We write the general \sph\ metric in \ME\ in the form
\beq                                                         \label{ds1}
    ds_E^2 = \og\mn dx^{\mu}dx^{\nu}
           = \e^{2\gamma} dt^2 - \e^{2\alpha}du^2 - \e^{2\beta}d\Omega^2,
\eeq
  where $\alpha$, $\beta$ $\gamma$ are functions of the unspecified radial coordinate $u$ 
  and the time coordinate $t$, while $d\Omega^2= d\theta^2 + \sin^2\theta\,d\varphi^2$,
  and the coefficient at this metric on a unit sphere, $\e^{2\beta} \equiv r^2(u,t)$
  is the squared spherical radius.  By definition, a center (if any) corresponds to $r \to 0$.
  
  The general \ssph\ solution of the theory \rf{S-GR} with the metric \rf{ds1} and an electric 
  field is well known since the 70-s \cite{penney, br73} and consists of a few branches, depending 
  on whether the field $\psi$ is canonical or phantom and on the integration constants, see the
  corresponding classification, e.g., in our previous paper \cite{we24}. In the present study, 
  being interested in solutions admitting a conformal continuation, we have to focus on 
  the single branch [1+] with a canonical scalar ($\eps = +1$). 
  In terms of the harmonic radial coordinate $u$, with which the metric coefficients in 
  \rf{ds1} satisfy the condition $\alpha = 2\beta + \gamma$, this solution has the form
\bear
     ds_E^2 \eql                                                  \label{ds_E}
         \frac{h^2 dt^2}{q^2 \sinh^2 [h(u+u_1)} -
     				 \frac{k^2 q^2  \sinh^2 [h(u+u_1)]}{h^2 \sinh^2(ku)}
			          \biggr[\frac{k^2 du^2}{\sinh^2(ku)} + d\Omega^2\biggl],
\yy
     \psi \eql  Cu,                                         \label{psi}
\\						\label{F_mn}
     F\mn \eql  \frac{h^2(\delta_{\mu 0}\delta_{\nu 1}
	    -\delta_{\nu 0}\delta_{\mu 1})}{q\,\sinh^2[h(u+u_1)]}, 
	    \ \then \ F\mn F\MN = - \frac{2q^2}{r^4} 
	    = - \frac{2 h^4 \sinh^4(ku)}{q^2 k^4 \sinh^4 [h(u+u_1)]},
\ear
  where $q$ (the electric charge\footnote
			{In addition to $q$, we might introduce a monopole magnetic charge $\oq$ 
		       instead of or in addition to $q$. This would not change any results of this study,
		       the only change in this more general case would be a replacement  
			 of $q$ with $\sqrt{q^2 + \oq^2}$ in all relations.}),
  $C$ (the scalar charge), $k >0, h>0$ and $u_1 > 0$ are 
  integration constants, three of them related by the equality
\beq                       \label{int}
		k^2 = h^2 + C^2.
\eeq  
  The coordinate $u$ is defined in the range $u>0$, such that the value $u =0$ corresponds to flat
  spatial infinity, while $u \to \infty$ is a central naked singularity where $r \to 0$ 
  (because $k >h$). The additional requirement on $u_1$
\beq              \label{u1}
		\sinh^2 (hu_1) = h^2/q^2
\eeq  
  provides $g_{00}\big|_{u=0} = 1$, that is, a natural choice of the time unit at spatial infinity. 
  Thus at small $u$ the conventional flat-space spherical radial coordinate $r=\e^{\beta}$ 
  is simply $r = 1/u$.

  Three essential integration constants of the solution are the charges $q$ and $C$ and 
  either $k$ or $h$. Moreover, comparing the asymptotic expression for $g_{00}$ in \rf{ds_E} 
  at small $u \approx 1/r$ with the Schwarzschild metric, we obtain the value  of the  
  \Scw\ mass $m$ of this space-time as 
\beq
	       m = \sqrt{q^2 + h^2},      		\label{m}
\eeq
  thus the solution is completely determined by the mass and two charges.

  In the absence of the \emag\ field ($q=0$), we deal with Fisher's scalar-vacuum solution 
  \cite{fisher} where the metric in terms of the harmonic coordinate $u$ has the form 
\beq                                                 \label{ds_E0}
     ds_E^2 = \e^{-2hu} dt^2 -	 \frac{k^2 \e^{2hu}}{\sinh^2(ku)}
			          \biggr[\frac{k^2 du^2}{\sinh^2(ku)} + d\Omega^2\biggl],  
\eeq  
  the scalar field is again $\psi = Cu$, the relation \rf{int} is also valid, and the \Scw\ mass  
  is simply $m = h$. This solution and its phantom counterpart have been well studied, 
  see, in particular, \cite{h-ell73, br73, kb-hod, br-book, sarb1, stab11, stab18}, and the stability 
  of the corresponding STT solutions with $\eps=+1$ was recently discussed in \cite{we23}.

\subsection{Brans-Dicke electrovacuum. Exceptional \whs.} 

  As already mentioned, the solution \rf{ds_E}--\rf{F_mn} does not only belong to GR but is 
  also a solution of an arbitrary STT in its Einstein frame \ME. The corresponding solution of 
  the Brans-Dicke theory in \MJ\ is obtained according to \rf{map} with  
  $\phi = \e^{\sqrt 2 \psi/\oom}$ (assuming $\psi_0 = 0$ without loss of generality).
  The \emag\ field $F\mn$ looks the same in both frames due to the conformal invariance of
  the \emag\ Lagrangian equal to $-\sqrt{-g}F\mn F\MN$. Thus the metric in \MJ\ is
 \beq            \label{dsJ-u}
		ds_J^2 = \e^{-\sqrt 2 Cu/\oom} \bigg[ \frac{h^2 dt^2} {q^2 \sinh^2[h(u+u_1)]}
					- \frac {4 q^2 \sinh^2[h(u+u_1)]} {\sinh^2 (2hu)}
       				\bigg(\frac{4 h^2 du^2}{\sinh^2 (2hu)} + d\Omega^2\bigg) \bigg],
\eeq    

  Our interest is now in a conformal continuation due to the transition \rf{map} from \ME\ to \MJ; 
  it becomes possible only if the quantities $\e^{2\gamma}$ and $\e^{2\beta} \equiv r^2$ 
  in the metric \rf{ds1}) of \ME\ vanish or blow up in the same manner, thus admitting a
  simultaneous ``correction'' by a suitable conformal factor. In the solution under study this 
  only happens under the condition 
\beq  
         k=2h  \ \then     C^2 = 3h^2 = 3(m^2 - q^2),  
\eeq         
  so that $\e^{2\gamma} \sim \e^{2\beta} \sim \e^{-2 hu}$ as $u\to \infty$.
  Furthermore, to get rid of the singularity, the conformal factor must behave in a precisely 
  opposite way, that is, $1/\phi \sim \e^{-2 hu}$ as $u\to \infty$. As is easily verified, it is the 
  case only with the coupling constant $\omega=0$ and also $C = -\sqrt 3 h$, when we simply 
  obtain $1/\phi = \e^{2hu}$. 

  Now, since in \rf{dsJ-u} $\e^{-\sqrt 2 Cu/\oom} = \e^{2hu}$, the metric is regular on the sphere 
  $u = \infty$, and therefore the space-time \MJ\ should be continued beyond it using a new 
  radial coordinate that takes a finite value at $u=\infty$. A convenient choice is 
  $y = \e^{-2hu}$. We then obtain
\beq         \label{BD-wh}
	  ds_J^2 = g\mn dx^\mu dx^\nu 
	  	= \frac{4 h^2 dt^2}{[m+h - y(m-h)]^2} 
	  	  - \frac {4 [m+h - y(m-h)]^2}{(1-y^2)^2}
		  \bigg( \frac{4 dy^2}{(1-y^2)^2} + d\Omega^2 \bigg),
\eeq	
  where $y \in (-1, 1)$. Let us note that $m = \sqrt{h^2 + q^2}$ is the \Scw\ mass in \ME\ 
  (recall \rf{m}), but in \MJ\ the mass has another value due to the conformal factor,
   and in the present case it is\footnote
   			{For a general metric of the form \rf{ds1} \asflat\ at some $u = u_0$, the 
   			\Scw\ mass can be obtained as the limit \cite{br-book}
   			$m_{\rm Sch} = \lim\limits_{u\to u_0} \gamma'\e^\beta/\beta'$.} 
   $m_J = \sqrt{h^2 + q^2} - h = m - h$.
  The original spatial infinity $u=0$ corresponds to $y=1$, while $y = -1$ is another spatial 
  infinity, and all metric coefficients are finite in the range $y^2 < 1$. 
  Thus the space-time \MJ\ is a static traversable \wh. It is \asflat\ at both infinities,
  but with different time rates at the two ends since 
\beq  
  		g^J_{00}\Big|_{y=1} = 1, \qq  
  		g^J_{00}\Big|_{y=-1} = \frac{h^2}{m^2} = \frac{h^2}{h^2 + q^2}. 
\eeq
  The \Scw\ masses are also different: at $y=1$ we have (as mentioned) $m_J = m_{J+} = m-h >0$,
  while at $y =-1$ we obtain $m_J = m_{J-} = h-m < 0$. 

  The \wh\ throat as a minimum of the spherical radius $r_J = \sqrt{-g_{22}}$ is located where 
  $d r_J/dy =0$, for which we find 
\beq
           y = y_{\rm th} = \frac{\sqrt m - \sqrt h} {\sqrt m + \sqrt h} > 0,
\eeq     
  and the throat radius is 
\beq
		r_{\rm th} = \min r_J(y) = \big[\sqrt m + \sqrt h\big]^2.		
\eeq    
  Thus at $q \ne 0$ (when $m > h$) we obtain a \wh\ asymmetric with respect to its throat,
  see Fig.\,1a for the profile of the spherical radius $r_J(y)$ at different $h$, assuming $m=1$..
  The solution depends on two integration constants $m$ and $h$ or, alternatively, 
  $m$ and $q = \pm\sqrt{m^2 - h^2}$. 
\begin{figure*}
\centering
\includegraphics[width=5.7cm]{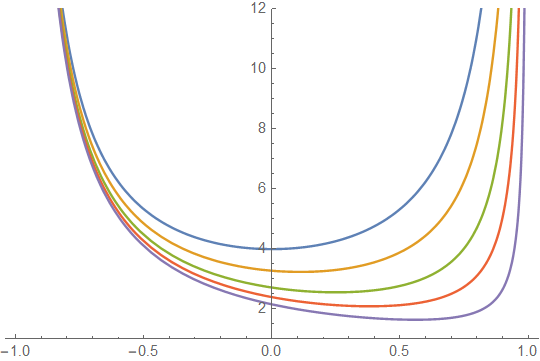}\ \
\includegraphics[width=5.7cm]{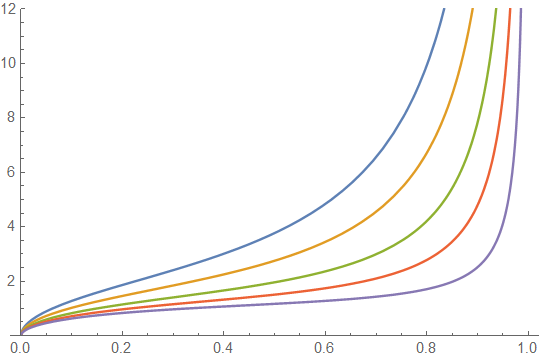}\ \
\includegraphics[width=5.7cm]{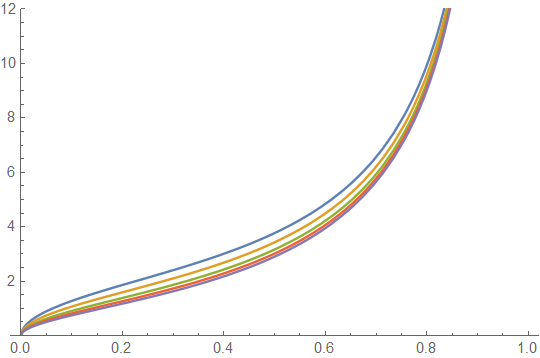}
\caption{\small
	The spherical radius vs.\,$y$ in \MJ\ and \ME. \\
	{\bf Left:} The radius $r_J(y),\ y\in (-1, 1)$ at different values of $h =\sqrt{m^2 - q^2}$, 
	assuming $m=1$ (upside down: $h = 1, 0.64, 0.36, 0.2, 0.08$). Each minimum of $r_J(y)$ is 
	a \wh\ throat.\\
	{\bf Middle:} The radius $r(y)$ in $\ME_+$ ($y>0$) for the same values of $m$ and $h$.\\
	{\bf Right:} The radius $r(|y|)$ in $\ME_-$ ($y<0$) for the same values of $m$ and $h$.
	Quite naturally, there are no \wh\ throats (minima of $r$) in $\ME_\pm$. 
	}  \label{r(y)}
\end{figure*}
  
  At $q =0$, the metric \rf{BD-wh} becomes symmetric and acquires an especially simple form if 
  the spherical radius $r_J = 4h/(1-y^2)$ is used as a new coordinate:
\beq
		ds_J^2\Big|_{q=0} = dt^2 - \frac{16 h^2}{(1-y^2)^2} 
								\bigg(\frac{4 dy^2}{(1-y^2)^2} + d\Omega^2\bigg)
				= dt^2 - \frac{d r_J^2}{1 - 4h/r_J}  - r_J^2 d\Omega^2. 
\eeq      
  The throat is then located at $y =0$, corresponding to $r_J = 4h$. This metric is sometimes 
  called ``a \Scw\ \wh'' since its spatial part coincides with that of the \Scw\ metric. 
  
  For $q \ne 0$, an attempt to introduce such a coordinate leads to rather inconvenient 
  expressions, so the description in terms of $y$ looks optimum.
  
\medskip
{\bf Remark.}
    It is known that in BD solutions belonging to the canonical part of the theory ($\omega > -3/2$)
    throats can exist at any $\omega$, even very large ones \cite{skvo10}. However, in all such cases 
    \whs\ as global configurations with good asymptotic behavior on both sides from the throat are
    impossible  \cite{br-star07}. The presently discussed solutions are the only examples of BD 
    \whs\ with $\omega > -3/2$. 

\medskip
\def\oy{{\overline y}}  
  
  The BD scalar field in the present solution with any $q$ is simply $\phi = \e^{2\psi/\sqrt 3} = y $, 
  and its sign changes at the transition through the regular sphere $y=0$. As in other cases of
  conformal continuations \cite{kb-CC1, kb-CC2}, the effective gravitational constant, 
  which is proportional to $1/\phi$, is negative beyond the transition surface $y =0$..
	
  The region $y >0$ of the \wh\ space-time \rf{BD-wh} corresponds to the whole
  Einstein-frame manifold \ME\ with the metric given by the expression in square brackets in 
  \rf{dsJ-u}, in which the singularity at $u=\infty$ corresponds to $y=0$, while it is a regular sphere 
  in \MJ. It is of interest to see what is the E-frame metric corresponding to the region 
  $y < 0$. So, assuming $y < 0$, let us for convenience denote $- y = \oy >0$ and substitute
  $\oy = \e^{-2hu}$, in full similarity with the transition from \rf{dsJ-u} to \rf{BD-wh}.
  Instead of \rf{dsJ-u}, we now obtain    
\beq            \label{dsJ-}
		ds_J^2 = \e^{2hu}\bigg[ \frac{h^2 dt^2} {q^2 \cosh^2[h(u+u_2)]}
					- \frac {4 q^2 \cosh^2[h(u+u_2)]} {\sinh^2 (2hu)}
					\bigg(\frac{4 h^2 du^2}{\sinh^2 (2hu)} + d\Omega^2\bigg) \bigg],
\eeq    
  where, again, $u \in \R_+$, and $u =0$ corresponds to spatial infinity. The constant $u_2$
  is determined by the relation 
\beq
			\cosh^2 (hu_2) = m^2/q^2.
\eeq  
  The metric in \ME\ corresponding to \rf{dsJ-}, that is, $ds^2_E = \oy ds_J^2$, is not a 
  solution to the Einstein-Maxwell-scalar equations, as was \rf{ds_E}, but it becomes such a  
  solution if we replace $q^2 \to -q^2$ in the \emag\ SET. An evident explanation of this 
  fact is that when the $\phi$ field changes its sign --- as happens at a transition to $y < 0$
  --- the gravitational field action also changes its sign, and it involves both the metric and 
  the scalar field $\phi$, hence also $\psi$ that emerges in the Einstein frame. Meanwhile,
  the \emag\ field action remains the same, therefore, the field equations now look as if 
  all \emag\ field contributions were multiplied by $-1$.  
  
  In the case $q=0$ ($h =1$), when the \wh\ is symmetric, the Einstein frame manifolds $\ME_+$ and 
  $\ME_-$, corresponding to the parts $y >0$ and $y< 0$ of the Jordan-frame manifold \ME, are
  identical. Unlike that, at $q \ne 0$ ($h < 1$), the geometries of $\ME_+$ and $\ME_-$ are different,
  as is illustrated by the behavior of $r(y)$ in Figs.\,1b and 1c.     
 
\section{Perturbation equations}  
  
  Let us now consider \sph\ perturbations: of the charged \wh\ solution. As in \cite{we23, we24},
  we will use its \ME\ representation as a tool since the \pb\ equations look much simpler in 
  the \ME\ variables. However, we have to deal now with two manifolds $\ME_+$ ($y > 0$)
  and $\ME_-$ ($y < 0$). Let us begin with $\ME_+$ and, instead of $\psi (u)$
  consider a perturbed function
\[
    \psi(u,t)= \psi(u)+ \dpsi(u,t)
\]
  and introduce in a similar way \pbs\ of the metric functions $\da,\ \db,\ \dg$
  in terms of the metric \rf{ds1}. 
  As in all such cases, the only dynamic degree of freedom is related to $\dpsi$ since the 
  gravitational and \emag\ perturbations cannot be purely radial (monopole). 
  Accordingly, using the perturbation gauge $\db \equiv 0$,\footnote
  		{It has been shown \cite{stab11, stab18, br-book} that the resulting wave equation 
  		  is gauge-invariant and thus describes real perturbations of the system rather than 
  		  pure coordinate effects. }
  quite similarly to \cite{kb-hod, stab11, stab18, br-book, br-kor15}, with the aid of the Einstein 
  equations we exclude the metric \pbs\ from the perturbed scalar field equation $\Box \psi =0$ 
  and separate the variables assuming
\beq  
 			\dpsi = \e^{i\Omega t} X(u),\qq \Omega = \const,
\eeq    
  to obtain the following equation for $X(u)$ written in terms of an arbitrary radial coordinate $u$:
\bearr  			\label{eq-X}
            X'' + (\gamma'+ 2\beta'-\alpha') X' 
        					+ [\e^{2\alpha-2\gamma}\Omega^2 - W(u)] X =0,
\yyy                     \label{W}
      W(u) \equiv  \frac{2 \psi'{}^2}{\beta'^2}\e^{2\alpha -2\beta}\big(q^2 \e^{-2\beta} -1\big)
		    \equiv  \frac{2 \e^{2\alpha} \psi'{}^2}{r'{}^2}\Big(\frac{q^2}{r^2} -1\Big).
\ear 
  A further substitution $X(u) = \e^{- \beta} Y(z)$, where $z$ is the ``tortoise'' coordinate 
  (obtained as $z = \int \e^{\alpha(u) -\gamma(u)} du$), while $\beta = \log r$ is taken from 
  the static background metric, leads to the standard \Schr-like form of the \pb\ equation 
  for $Y(z)$ \cite{kb-hod, stab18}:
\beq                                                        \label{Schr}
      \frac {d^2 Y}{dz^2} + [\Omega^2 - \Veff(z)] Y =0.
\eeq
  Here the effective potential $\Veff$ is expressed, again in terms of an arbitrary coordinate 
  $u$, as
\beq  			\label{Veff-u}
  	\Veff(u) = \e^{2\gamma-2\alpha}\big[W(u)+ \beta'' + \beta'(\beta'+\gamma'-\alpha')\big].
\eeq  

  Since all these relations are valid in \ME, when applying them to our \wh\ solution, we have 
  to use them twice, once for $\ME_+$ and once for $\ME_-$, using the corresponding 
  metrics $ds_E^2$. However, since the \wh\ \rf{BD-wh} must be considered as a unified 
  physical system, its viable \pbs\ are those with the ``frequencies'' $\Omega$ common for 
  $\ME_+$ and $\ME_-$. The important question on boundary conditions at their separating
  surface $y =0$ will be discussed below.
  
  Let us begin with $\ME_+$ ($y > 0$). We have to notice that, as in many previous papers
  such as \cite{kb-hod, we23}), the coordinates $u$ or $y$ used in the solution under study
  cannot be expressed analytically in terms of the tortoise coordinate $z(u)$, making it
  impossible to present $\Veff(z)$ as an explicit function. Therefore, it makes sense to use 
  \eqn{Schr} for asymptotic analysis and possible qualitative inferences, but it cannot be 
  solved exactly, while a numerical analysis is more reasonable with \eqn{eq-X}, 
  written in terms of the coordinate $y$ used in our background solution. 
  
  The metric in in $\ME_+$ has the form \rf{ds_E} with $k=2h$ in terms of the harmonic 
  coordinate $u$, or, if expressed in terms of $y$, it is \rf{BD-wh} multiplied by $y$, see the 
  behavior of the spherical radius $r = r_J$ in Fig.\,\ref{r(y)}, right panel. 
  There is a naked central singularity at $y=0$, where $r =0$, and a spatial infinity at $y=1$ 
  with the \Scw\ mass $m$. 
 
  The functions $W$ and $\Veff$ involved in the \pb\ equations are  
\bearr                              \label{W+}
		W_+(y) =  \frac{6 m^2 (1 - y)^4 - 6 h^2 (1 + y)^4 - 48 h m y (1 - y^2)}
				{y [m (1 - y)^3 + h (1 + y)^3]^2},
\yyy                             \label{Veff+}
             \Veff{}_+ (y) = - \frac {h^2 (1 - y^2)^3}
             {16 y^2 [h + m + (h  - m) y]^6 [m (1 - y)^3 + h (1 + y)^3]^2} 
\nnn \qq \times
		\Big[m^4 (1 - y)^7 (-1 - 25 y + 25 y^2 + y^3) - h^4 (1 + y)^7 (-1 - 25 y + 25 y^2 + y^3) 
\nnn \cm
		 +  6 h^2 m^2 (1 - y^2)^3 (1 + 70 y^2 + y^4)
\nnn \cm
		+ 4 h m^3 (1 - y)^4 (1 - 12 y + 39 y^2 + 136 y^3 + 39 y^4 - 12 y^5 +  y^6)
\nnn   \cm 
	    + 4 h^3 m (1 + y)^4 (1 + 12 y + 39 y^2 - 136 y^3 + 39 y^4 + 12 y^5 +  y^6)\Big].
\ear    
  
  Let us consider the asymptotic properties of $X(y)$ and $Y(z)$.
  At spatial infinity $y\to 1$, we have 
\beq                  \label{as1+}      
		z(y) \approx r(y) \approx \frac{2h}{1-y}, \qq  
		y \approx 1 -\frac{2h}{z}, \qq
		\Veff \approx \frac{2m}{z^3}.		
\eeq  
  Then the approximate behavior of solutions to \eqn{Schr} is
\beq               \label{Y0+}
			Y(z) \approx C_1 \e^{|\Omega|z} + C_2 \e^{-|\Omega|z}
\eeq
  under the assumption $\Omega^2 <0$ (as occurs at an instability with exponential 
  growth of \pbs), and
\beq               \label{Y00}
			Y(z) \approx C_3 + C_4 z
\eeq  
  assuming $\Omega = 0$ (with a possible linear growth of \pbs, $\dpsi \sim t$). 
  Here and henceforth $C_i$ denote integration constants. 
   
  At the singularity $y=0$ we can put $z = 0$, and in its neighborhood
\beq                   \label{as0}
 		z \approx \frac{2(m+h)^2}{h} y, \qq 
 		y \approx \frac{hz}{2(m+h)^2}, \qq
 		\Veff \approx - \frac{1}{4z^2}.
\eeq 
  The solution to \eqn{Schr} near $z=0$ with any $\Omega$ has the form
\beq 				\label{Y0}
		Y(z) \approx \sqrt z (C_5 + C_6 \log z), \qq z\to 0, \qq C_5, C_6 = \const.
\eeq  

   What shall we find in the space-time $\ME_-$ conformal to the region $y <0$ in \MJ?
   Let us, for convenience and without risk of confusion, write again $y$ instead of $-y$ or $|y|$.
   
   The metric has the form given within the square brackets in \rf{ds_J-}, or in terms of $y$,
\beq         \label{ds_E-}
	  ds_E^2 = \og\mn dx^\mu dx^\nu 
	  	= \frac{4 h^2 y\, dt^2}{[m+h + y(m-h)]^2} 
	  	  - \frac {4 y [m+h + y(m-h)]^2}{(1-y^2)^2}
		  \bigg( \frac{4 dy^2}{(1-y^2)^2} + d\Omega^2 \bigg),
\eeq	   
   Like its counterpart in $\ME+$, this metric has a naked central singularity at $y=0$ and a 
   spatial infinity at $y=1$, but now the \Scw\ mass is equal to $h$. The functions $W$ and 
   $\Veff$ are slightly different from \rf{W+} and \rf{Veff+}:
\bearr                              \label{W-}
		W_-(y) =  \frac{6 h^2 (1 - y)^4 - 6 m^2 (1 + y)^4 - 48 h m y (1 - y^2)}
				{y [h (1 - y)^3 + m (1 + y)^3]^2},
\yyy                             \label{Veff-}
             \Veff{}_- (y) = -  \frac {h^2 (1 - y^2)^3}
             {16 y^2 [h + m + (m - h) y]^6 [h (1 - y)^3 + m (1 + y)^3]^2} 
\nnn \qq \times
		\Big[ h^4 (1 - y)^7 (1 - 25 y - 25 y^2 + y^3) - m^4 (1 + y)^7 (-1 - 25 y + 25 y^2 + y^3) 
\nnn \cm
		 + 6 h^2 m^2 (1 - y^2)^3 (1 + 70 y^2 + y^4)
\nnn \cm
		 + 4 h^3 m (1 - y)^4 (1 - 12 y + 39 y^2 + 136 y^3 + 39 y^4 - 12 y^5 + y^6)
\nnn   \cm 
	       + 4 h m^3 (1 + y)^4 (1 + 12 y + 39 y^2 - 136 y^3 + 39 y^4 + 12 y^5 + y^6)\Big].
\ear     
 
   At spatial infinity $y\to 1$, we now have 
\beq 			\label{as1-}
		z(y) \approx \frac mh r(y) \approx \frac{2m^2}{h (1-y)}, \qq  
		y \approx 1 -\frac{2m^2}{h z}, \qq
		\Veff \approx \frac{2m}{z^3}.		
\eeq  
  Since here again $\Veff$ rapidly vanishes as $z\to\infty$, we have the same 
  solutions to \eqn{Schr} given by \rf{Y0+} and \rf{Y00}.

  Near the singularity $y=0$, putting there again $z = 0$, we obtain the same
  asymptotic behavior \rf{as0} as in $\ME_+$ and consequently the same asymptotic 
  solution \rf{Y0} to \eqn{Schr}.
 
  Thus the asymptotic properties of \pbs\ are similar in $\ME_+$ ($y>0$) and 
  $\ME_-$ ($y< 0$), but the effective potentials are different, as illustrated in Fig.\,2.
\begin{figure*}
\centering
\includegraphics[width=7.3cm]{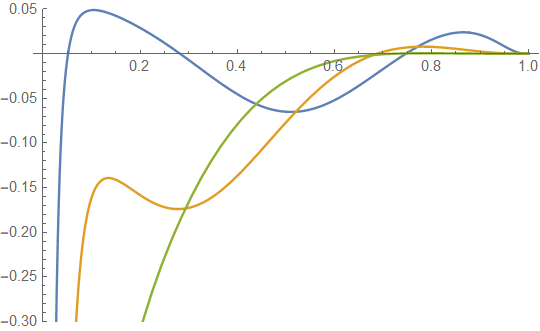}\qq
\includegraphics[width=7.3cm]{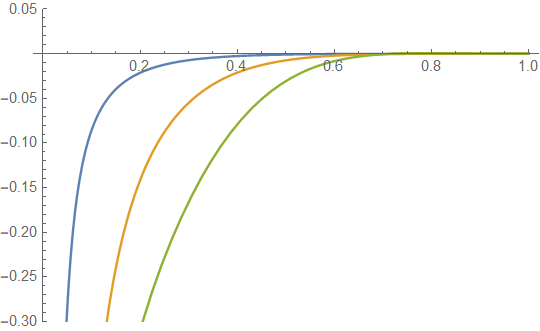}
\caption{\small
	The effective potentials for \pbs\ $\Veff$ vs. $|y|$ in space-times $\ME_+$ ($y> 0$ --- left panel) 
	and $\ME_-$ ($y< 0$ --- right panel) for $m=1$ and $h = 0.1, 0.35, 1$.  
	The value $h=1$ corresponds to $q=0$, in which case the two potentials are identical. }   \label{Veff}
\end{figure*}
  
\section{Boundary conditions and stability}  
\subsection{Boundary conditions} 
  
  To study the stability of our static background configuration, we must seek solutions to
  \eqs \rf{eq-X} or \rf{Schr} satisfying physically meaningful boundary conditions 
  (assuming, in particular, finite \pb\ energy and absence of ingoing waves) and determine 
  the eigenvalues $\Omega^2$ admitting such solutions. If there are eigenvalues 
  $\Omega^2\leq 0$, we can conclude that the background solution is unstable since the
   perturbation $\dpsi$ can grow with time exponentially (if $\Omega^2 < 0$) or linearly
   if $\Omega =0$), and if such solutions are proved to be absent, we conclude that
  the background system is linearly stable under this kind of perturbations.
  
  In the system under study, the Jordan frame \MJ\ is physically preferred, and we must 
  formulate the boundary conditions in this frame, even though \eqs \rf{eq-X} or \rf{Schr}
  are written using variables belonging to \ME, obtained from \MJ\ by the substitutions \rf{map}. 
  The conditions must be imposed at the two spatial infinities $y = \pm 1$ and at the 
  surface $y =0$ that makes a singular boundary both in $\ME_+$ and $\ME_-$. 
  A universal requirement for scalar field \pbs\ used in \cite{kb-hod, we23, we24}, used even 
  at singularities, is $|\df/\phi| < \infty$, meaning that even if the background field blows up 
  somewhere, the \pb\ must not grow faster. In our system, the relevant scalar field in \MJ\
  is the everywhere finite field $\phi = y$. However, a stronger requirement 
  $\df \lesssim 1/r \sim 1- |y|$ as $y \to \pm 1$ follows from the condition of a finite energy 
  of \pbs\ that leads to $|\dpsi| \lesssim 1- |y|$ at these both boundaries, which in turn leads 
  to the conditions for the functions $X(y)$ and $Y(z)$ used in \eqs \rf{eq-X} and \rf{Schr}:
\beq               \label{bc-infty}
		  |X|/(1-|y|) \sim |X| z < \infty, \qq  |Y| < \infty\ \ \ {\rm as}\ \ \ y\to \pm 1,
\eeq
  since $Y(z) = r X(y) \approx z X(y)$ at large $z$. Moreover, as follows from \rf{Y0+}, this 
  condition leads to $C_1=0$ and $Y\to 0$ as $z \to \infty$ for $\Omega^2 < 0$, and 
  only at $\Omega =0$ a finite $Y$ is admitted at $z\to\infty$ ($|y|\to 1$) while $C_4 =0$.
  
  A more subtle reasoning is required at $y =0$. We might quite formally apply the 
  condition $|\df/\phi| < \infty$ and, since $\phi =y$, obtain the requirement 
  $|\df/y| < \infty$. However, then it would remain unclear why we forbid finite \pbs\ of 
  the field $\phi$. Still let us recall that the effective gravitational constant $G_{\rm eff} \sim 1/\phi$
  (see, e.g., \cite{BD-STT}), it blows up at $y=0$, and it seems quite reasonable to forbid 
  its perturbations blowing up even faster, that is, we should require 
  $|\delta G_{\rm eff}/G_{\rm eff}| = |\df/\phi| < \infty$, and this in turn leads 
  to $|\df/y| < \infty$. One can also verify that the same condition $ |\df/\phi| < \infty$ provides
  finite values for \pbs\ of the metric coefficients in \MJ\ at the regular surface $y=0$, which 
  should evidently be the case. In other words, this boundary condition provides unity of the 
  two halves of \MJ\ when subject to \pbs.
  
  For the functions $X(y)$ and $Y(z)$ we thus obtain the conditions
\beq  \wide
		|X(y)| < \infty, \qq  |Y(z)|/\sqrt z < \infty \ \ \ {\rm as}\ \  y \to 0 \ \ {\rm and}\ \ z \to 0,    
\eeq  
  since $Y(z) = r X(y)$, and $r \sim \sqrt {|y|} \sim \sqrt {z}$ at small $|y|$.
  Then, in the asymptotic solution \rf{Y0} we should require $C_6 =0$.
  (We would here remind the reader that $r(y)$ is the radius in \ME\ that vanishes at $y=0$, 
  while in \MJ\ the corresponding quantity $r_J(y)$ is finite.)
  
  As a result, in each of the manifolds $\ME_\pm$, we have asymptotic solutions to \eqn{Schr}
  at each end of the range of $z$ ($z \in \R_+$), where our boundary conditions select one 
  of two linearly independent solutions. This leads to well-posed boundary-value problems  
  in $\ME_\pm$.
    
\subsection{Numerical analysis} 
  
  We have mentioned that since the function $W$ in \eqn{eq-X} and the effective potential 
  $\Veff$ are expressed in terms of $y$ instead of $z$, it makes sense to study numerically 
  \eqn{eq-X} with boundary conditions formulated for $X(y)$.
  
  Thus in both $\ME_+$ and $\ME_-$ we consider \eqn{eq-X} in the form
\beq  			\label{eq-X1} 
            X'' + \frac{X'}{y} 
            + \bigg[\frac{4 [m + h \mp y (m-h)]^4}{h^2(1-y^2)^4}\Omega^2 - W_\pm(y)\bigg] X =0,
\eeq            
   where $W_\pm (y)$ are given by \eqs \rf{W+} and \rf{W-}. 
   Note that for a convenient comparison of the two, we replace $-y = |y|$ in $\ME_-$ with $y$. 
   Curiously, the expressions \rf{W+} and \rf{W-} differ from each other by the replacement
   $m \leftrightarrows h$.
   
   With \eqn{eq-X1}, we have the boundary conditions $X(0) = \const \ne 0$ and $X(1) =0$
   for $\Omega^2 < 0$ and $|X(1)| < \infty$ for $\Omega =0$.
   
   In the numerical shooting method, it is impossible to place the initial point at $y=0$ since it 
   is a singular point of \eqn{eq-X1}, but we can take such a point at some $y_0 \ll 1$,
   to impose there the conditions $X(y_0) = X_0 >0$ and $X'(y_0) = 0$, and to solve the 
   equation numerically in order to find such values of $\Omega$ that lead to suitable $X(1)$.  
   All that must be done separately for $\ME_+$ and $\ME_-$.
   
   We implement the Runge-Kutta procedure for solving \eqn{eq-X1} with the boundary conditions
   specified above. The variable $y$ ranges in the interval $(y_0, y_1)\sim(0.001, 0.999)$ corresponding 
   to the appropriate numerical accuracy. Without loss of generality, we put $m=1$, which fixes the 
   length scale, and $X_0=1$, which particular value is insignificant since \eqn{eq-X1} is linear.
   
   In the framework of the shooting method, we solve \eqn{eq-X1} with the initial conditions 
   $X(y_0) = 1$ and $X'(y_0) = 0$, separately for $\ME_+$ and $\ME_-$, with some test negative 
   value of $\Omega^2$, and obtain the corresponding numerical solution $X_{\rm num}(y; \Omega)$. 
   If the chosen value of $\Omega^2$ is not an eigenvalue of our problem, the curve 
   $X_{\rm num}(y; \Omega)$ strongly blows up on the right end $y_1$, whereas in the case of an 
   eigenvalue the numerical curve tends to a small value at $y_1$. Therefore, tracking the behavior 
   of the curves $X_{\rm num}(y; \Omega)$ at the right end, we find an eigenvalue $\Omega^2$ 
   (if any) and reveal the corresponding instability regions for different values of the free parameters 
   of the system. In our case, after fixing $X_0$ and $m$, there is just one free parameter 
   $h\in(0, m]=(0, 1]$.

\begin{figure}    
\centering     
\includegraphics[width=7.5cm]{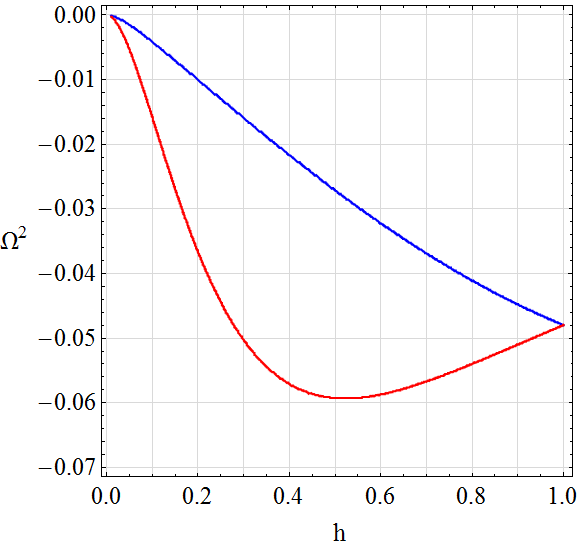}\\
\caption{\small 
		The eigenvalues $\Omega^2$ as functions of $h$. The top (blue) curve 
		corresponds to $\ME_+$, the bottom (red) one to $\ME_-$.
		}     \label{2omegas}
\end{figure}

\begin{figure}    
\centering     
\includegraphics[width=7.5cm]{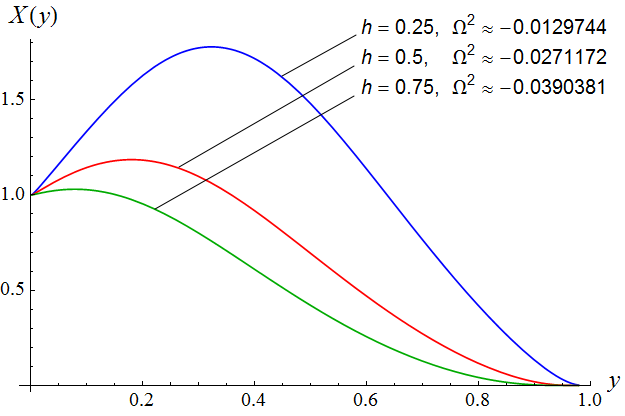}\qq
\includegraphics[width=7.5cm]{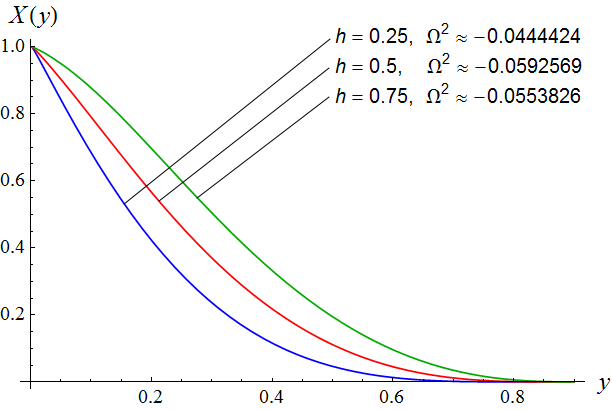}\
\caption{\small    
		The eigenfunctions $X(y)$ for some eigenvalues $\Omega^2$ obtained in $\ME_+$ ({\bf left}) 
		and in $\ME_-$ ({\bf right}).
		}   \label{eigen-f}
\end{figure}

   The results of our numerical analysis are presented in Fig.\,\ref{2omegas}. The plot  shows the
   existence of negative eigenvalues $\Omega^2$ as functions of $h$ for both manifolds $\ME_+$ 
   and $\ME_-$. Separately, in each of the two manifolds there are eigenvalues $\Omega^2$ in the 
   whole range of $h$ (some of theior eigenfunctions are shown in Fig.\,\ref{eigen-f}).
 . However, one can see that the eigenvalues in these two cases are everywhere
   different, except for the points $h=0$ (which does not belong to the solution range) and $h=1$
   (corresponding to $q = 0$). It means that the perturbations have no common spectrum with 
   $\Omega^2 \leq 0$ in $\ME_+$ and $\ME_-$,  hence no nonpositive eigenvalues in the entire 
   space-time \MJ\ for any $h\in (0, 1)$, that is, $q\ne 0$), but such a negative eigenvalue does 
   exist at $h =1$ ($q=0$). Thus our numerical analysis leads to a conclusion that charged \whs\ 
   under consideration are stable but electrically neutral ones are unstable. 

\section{Concluding remarks}    

   We have discussed the linear stability problem for an exceptional wormhole solutions of the 
   Brans-Dicke theory of gravity with the coupling constant $\omega=0$. In the Jordan frame it
   is rather hard to consider \pb\ equations, so, as in many studies including our previous ones
   \cite{we23, we24}, we used as a tool a transition to the Einstein frame, which, from the standpoint of
   differential equations, is simply a transition to other unknown functions. However, in the present 
   case the Jordan-frame manifold \MJ\ splits into two parts, each corresponding to its own 
   Einstein-frame manifolds $\ME_+$ or $\ME_-$, and the problem has to be solved in each of 
   them separately. Meanwhile, since \MJ\ is a unique smooth manifold, its \pbs\ should also be
   unified, which means that they must be finite and smooth everywhere in \MJ\ (in particular,
   on the boundary $y=0$), and the admissible modes must have common frequencies (be they 
   real or imaginary). Our study shows that such \pbs\ exponentially growing with time exist only
   for  electrically neutral \whs, and that growing modes are absent for charged ones. In other words,
   they are stable under linear monopole \pbs. 
   
   This study, in our opinion, gives an interesting example of a stabilizing role of transition surfaces
   at conformal continuations: such surfaces require certain boundary conditions, which leads to 
   solving different boundary-value problems ``to the left'' and ``to the right'' of them. 
   Another known example of such surfaces has been discovered when considering scalar fields  
   that admit transitions from canonical to phantom behavior (``trapped ghosts'')
   \cite{trap10, trap17}. It turns out that the transition surfaces where a scalar field changes its
   nature can be a regular surface in space-time, and physical requirement to the behavior of 
   \pbs\ on such surfaces are rather similar to those which we saw in this paper \cite{trap17, stab18},
    so these surfaces can also play a stabilizing role. An attractive feature of the trapped ghost concept
    is the opportunity to obtain \wh\ models where a scalar field is phantom only in a strong field 
    region and behaves as a usual canonical one outside it, as is favored by the experiment. 
    Construction and studies of such models of \whs\ (and probably other objects of interest) can 
    be a promising area of research.

\Funding
{The research of K. Bronnikov, S. Bolokhov and M. Skvortsova was supported by 
	RUDN University Project FSSF-2023-0003.
      F. Shaymanova and R. Ibadov gratefully acknowledge the support from
      Agency for Innovative Development under the Ministry of Higher Education, 
      Science and Innovation of the Republic of Uzbekistan, Project No. FZ-20200929385.
}   
        

\end{document}